%% file: hybrid.tex
\documentclass[sigconf]{acmart}

\usepackage{algorithm2e}
\SetKwFunction{saturate}{saturate}

\usepackage{subcaption}

\usepackage{multirow}

\AtBeginDocument{%
  \providecommand\BibTeX{{%
    \normalfont B\kern-0.5em{\scshape i\kern-0.25em b}\kern-0.8em\TeX}}}

\copyrightyear{2022}
\acmYear{2022}
\setcopyright{rightsretained}

\acmConference[IXR '22]{Proceedings of the 1st Workshop on Interactive eXtended Reality}{October 14, 2022}{Lisboa, Portugal}
\acmBooktitle{Proceedings of the 1st Workshop on Interactive eXtended Reality (IXR '22), Oct. 14, 2022, Lisboa, Portugal}
\acmDOI{10.1145/3552483.3556455}
\acmISBN{978-1-4503-9501-4/22/10}

\usepackage{etoolbox}
\makeatletter
\patchcmd{\maketitle}{\@copyrightpermission}{
   \begin{minipage}{0.3\columnwidth}
     \href{https://creativecommons.org/licenses/by/4.0/}{\includegraphics[width=0.90\textwidth]{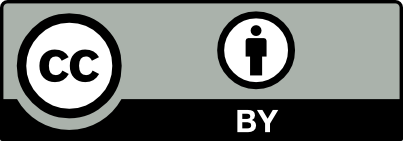}}
   \end{minipage}\hfill
   \begin{minipage}{0.7\columnwidth}
     \href{https://creativecommons.org/licenses/by/4.0/}{This work is licensed under a Creative Commons Attribution International 4.0 License.}
   \end{minipage}

   \vspace{5pt}
}{}{}

\makeatother

\acmSubmissionID{ixr008}

\begin{document}

\title[DHR]{DHR: Distributed Hybrid Rendering for Metaverse Experiences}

\author{Tan Yu Wei}
\affiliation{%
 \institution{National University of Singapore}
 \country{Singapore}
}
\email{yuwei@u.nus.edu}

\author{Alden Tan}
\affiliation{%
 \institution{National University of Singapore}
 \country{Singapore}
}
\email{alden.tan@u.nus.edu}

\author{Nicholas Nge}
\affiliation{%
 \institution{National University of Singapore}
 \country{Singapore}
}
\email{e0406660@u.nus.edu}

\author{Anand Bhojan}
\affiliation{%
 \institution{National University of Singapore}
 \country{Singapore}
}
\email{banand@comp.nus.edu.sg}

\renewcommand{\shortauthors}{Tan Yu Wei, Alden Tan, Nicholas Nge, \& Anand Bhojan}

\begin{abstract}
	Classically, rasterization techniques are performed for real-time rendering to meet the constraint of interactive frame rates. However, such techniques do not produce realistic results as compared to ray tracing approaches. Hence, hybrid rendering has emerged to improve the graphics fidelity of rasterization with ray tracing in real-time. We explore the approach of distributed rendering in incorporating real-time hybrid rendering into metaverse experiences for immersive graphics. In standalone extended reality (XR) devices, such ray tracing-enabled graphics is only feasible through pure cloud-based remote rendering systems that rely on low-latency networks to transmit real-time ray-traced data in response to interactive user input. Under high network latency conditions, remote rendering might not be able to maintain interactive frame rates for the client, adversely affecting the user experience. We adopt hybrid rendering via a distributed rendering approach by integrating ray tracing on powerful remote hardware with raster-based rendering on user access devices. With this hybrid approach, our technique can help standalone XR devices achieve ray tracing-incorporated graphics and maintain interactive frame rates even under high-latency conditions. 
\end{abstract}



\maketitle

\input{introduction}
\input{relatedwork}
\input{design}
\input{evaluation}
\input{futurework}
\input{conclusion}

\begin{acks}
	We thank Low Siang Ern for helping to improve the writing and quality of the paper. This work is supported by the Singapore Ministry of Education Academic Research grant T1 251RES2205, “Real-time Distributed Hybrid Rendering with 5G Edge Computing for Realistic Graphics in Mobile Games and Metaverse Applications”.
\end{acks}

\bibliographystyle{ACM-Reference-Format}
\balance
\bibliography{hybrid}

\end{document}

%% file: introduction.tex
\section{Introduction}

The metaverse refers to a collective virtual environment on the Internet that incorporates both the physical and digital world \citep{Lee:2018:AON} as empowered by extended reality (XR) technology for immersive real-time interaction and collaboration. According to virtual reality (VR) pioneer and metaverse expert Tony Parisi \citep{Parisi:2021:TSR}, access devices for the metaverse are not limited to standalone XR devices but can also include mobile phones and display walls. However, he acknowledges that one's experience of the metaverse can be enhanced with ``immersive hardware devices''. Hence, we target standalone XR devices to elevate the quality of the metaverse.

Ray tracing is a rendering technique which generates realistic lighting and camera effects by simulating the propagation of light with respect to scene geometry and the camera. However, it is computationally heavy due to the costs of calculating ray-geometry intersections and multiple shading per pixel. In contrast, rasterization provides a fast approximation of ray tracing's visual quality but suffers from certain inaccuracies which cannot be solved with just screen-space information. Given the advancements in hardware-accelerated personal computer (PC) graphics processing units (GPUs) that support ray tracing such as the NVIDIA GeForce RTX, we can now achieve more convincing graphics in real-time with a reasonable additional performance cost using ray tracing techniques that were once only feasible in offline rendering.

However, for standalone XR devices such as VR headsets, GPUs that support hardware-accelerated ray tracing are not feasible to realise with current technologies due to their size factor, heat generation and power consumption. Hence, such system hardware cannot perform real-time ray tracing fast enough to meet the requirements of 90 frames per second (FPS) for interactive XR applications. Nonetheless, ray tracing-based rendering can generate advanced lighting and camera effects for more photorealistic graphics as compared to traditional rasterization, helping to simulate a more immersive digital environment for the metaverse. Hybrid rendering for real-time applications seeks to combine ray tracing and rasterization techniques for a performance-quality tradeoff, producing higher fidelity results than classic real-time rendering while still maintaining interactive frame rates. However, on standalone XR devices, even hybrid rendering with only some ray tracing is not feasible to meet the tight performance constraint. 
 
In cloud gaming, remote rendering is often performed where the user's access device (the client) only handles the player's inputs and sends them to the remote server for rendering. After the rendering is done, output video frames are then streamed to the client to be displayed to the user. Hence, hybrid rendering can be performed as part of remote rendering which leverages the graphics capability of remote servers for rendering as compared to local rendering within the device itself. However, the cloud gaming industry is still facing multiple challenges including well-known latency and bandwidth issues. For instance, latency limits the proximity that a cloud server can support and in turn the number of clients, making cloud gaming not as attractive as other cloud-based services. Hence, the industry is focusing on cloud-assisted rather than cloud-dependent systems. 
 
We propose the novel cloud-assisted approach of hybrid rendering distributed between the cloud system and the client which we call Distributed Hybrid Rendering (DHR). DHR leverages both the client device and remote server for rendering by allocating the rendering workload based on real-time ray tracing capability. The client and server work together to generate ray tracing-incorporated high-fidelity graphics while maintaining interactive frame rates for metaverse experiences. By distributing the rendering workload, we can leverage the client devices' limited capability for rasterization to also serve as a backup should the ray-traced data from the remote server be delayed or lost. When this data is missing, the use of the application is not disrupted as local rasterization in the client with some approximation mechanisms can still provide desirable visual quality. DHR can be added directly to the rendering engine with no interference from the application developer. 

This paper is an extended version of our previous work \citet{Tan:CAH:2021} with the following key contributions:

\begin{itemize}
    \item Details of the design and implementation of a basic DHR prototype for dynamic ray-traced shadows.
    \item A frame prediction mechanism to maintain spatial detail and temporal coherence under adverse network conditions.
    \item Qualitative and quantitative evaluation of our technique, including the analysis of key factors such as network ping and the number of predicted frames.
\end{itemize}

%% file: relatedwork.tex
\section{Related Work}

Hybrid rendering combines ray tracing and rasterization to generate high-quality graphics while maintaining interactive frame rates for real-time applications. Instead of full ray tracing which might be too computationally expensive for geometrically complex scenes, ray tracing is typically at most partially employed such as for certain lighting effects \citep{Cabeleira:2010:CRR,Barre:2018:HRR} or pixels \citep{Marrs:2018:ATA,Macedo:2018:FRR,Beck:1981:CHR,Hertel:2009:HGR,Lau:2009:FHS} where the visual result of rasterization is less desirable. We employ hybrid rendering to produce ray tracing-enabled high-fidelity graphics for metaverse experiences while meeting interactive performance budgets.

Among the four main delivery modes for game contents \citep{Bhojan:2017:CTL}, video streaming and graphics streaming (or image-based streaming and instruction-based streaming as termed by \citet{Chan:2017:HMC}) are typically used with regards to rendering. Image-based streaming is where the server performs the rendering \citep{Holthe:2009:GLR}, and instruction-based streaming is where the server only computes the graphics commands for the rendering that the client proceeds to perform \citep{Eisert:2008:LDS}. Although instruction-based streaming reduces the amount of data that needs to be transmitted (i.e. graphics commands instead of whole video frames), image-based streaming is the convention for cloud gaming as rendering can be performed on powerful remote GPUs to achieve high-quality graphics even for thin-client user access devices. Popular cloud gaming services like Google Stadia \citep{Google:2021:S}, NVIDIA GeForce NOW \citep{NVIDIA:2021:GN} and Amazon Luna \citep{NVIDIA:2021:L}, all employ image-based streaming (i.e. remote rendering). Standalone XR devices are able to produce realistic visual results through the remote rendering of hybrid rendering techniques, provided that the network latency is low for the interactive streaming of video content to the client. However, under high-latency conditions, this pure cloud-based solution may not fulfil the requirement of interactive frame rates \citep{Bhojan:2014:GMM,Bhojan:2017:CTL} if data transmission is too slow, resulting in undesired lag for the user.

To reap the benefits of both the image- and instruction-based streaming modes, \citet{Chan:2017:HMC} propose a hybrid-streaming workflow that uses image-based streaming for far objects and instruction-based streaming for near objects, overlaying the results in order of depth. This workflow adopts the approach of collaborative or distributed rendering \citep{Cuervo:2015:KHM} where rendering is performed on both the client and server and their results are combined for the final output. Other distributed rendering techniques include layered coding \citep{Chuah:2014:LCM,Chuah:2016:LBM} where the server renders two versions of the frame of different quality based on polygon count and the complexity of lighting effects etc., while the client only renders the lower quality version. The pixel-wise colour difference between the versions is then sent to the client for recovering the higher quality result. The rendering workload can also be assigned in terms of the number of edges in scene models \citep{Chen:2019:FAR} and the number of frames to render \citep{Cuervo:2015:KHM}.

Similarly, we adopt a distributed rendering approach but for a hybrid rendering pipeline. Instead of performing both ray tracing and rasterization on the server, we offload ray tracing to the server while performing rasterization locally, combining the results. This also allows for better utilization of the client hardware as it can still perform rasterization locally and meet interactive frame rates.

In current cloud gaming services, the gameplay of the user is susceptible to the stability of the network. Lag and jitter can result which are detrimental to the interactiveness of metaverse experiences. This is especially crucial for standalone XR devices used on the go where network connections can be unstable as compared to in a controlled environment such as their home with a hardwired Ethernet connection or a fast wireless router. Nonetheless, we maintain a baseline of temporal coherence for the user by falling back on a lower quality result if the network connection is unstable. We obtain this result by predicting the data from the server if it does not arrive on time and combining it with local rasterization, generating approximate yet up-to-date interactive output for the user even under poor network conditions. As such, although our approach works best under fast and stable connections, it can also handle slow and unstable connections well unlike fully remote rendering. \citet{Cuervo:2015:KHM} allows both online and offline gaming. However, we handle variable network delay and jitter instead of just mid-game disconnections and provide an approximate prediction of the actual output instead of simply displaying the local render as-is. 

Our technique, \citet{Cuervo:2015:KHM}'s delta encoding approach, \citet{Chen:2019:FAR} as well as \citet{Chuah:2014:LCM} and \citet{Chuah:2016:LBM} can always locally render frames with geometrically accurate scene representation. The client already has all the data it needs to interactively display a reasonable low-quality output to the user in real-time. In contrast, for \citet{Chan:2017:HMC} and \citet{Cuervo:2015:KHM}'s client-side I-frame rendering technique where a subset of the scene geometry and frames respectively are only rendered at the server, scene representation as shown in the output will be outdated in the event of network delay and jitter. This same issue also exists for fully remote rendering where the client is only set up to display video output rather than perform any sort of rendering locally. We also save on bandwidth as compared to remote rendering and the distributed rendering techniques \citep{Cuervo:2015:KHM,Chen:2019:FAR,Chuah:2014:LCM,Chuah:2016:LBM} as we send variable-size visibility bitmaps based on the number of lights in the scene from the server as compared to streaming fixed-size full-colour buffers to the client. We test our approach on ray-traced shadows, but there are also global illumination approaches in the same vein \citep{Crassin:2015:CSA,Stengel:2021:DDS,Magro:2020:CBD} where indirect lighting is computed on servers and direct lighting is calculated on clients.

%% file: design.tex
\section{Design}

We perform ray tracing on remote servers, leveraging hardware-accelerated GPUs. Nonetheless, thin clients like standalone XR devices can still attain desirable performance with rasterization. Hence, we also leverage their limited graphics capability for rasterization. DHR can give us not only interactive frame rates through raster-based rendering on the client, but also high-quality graphics through ray tracing on the server. We study our distributed rendering approach on a simple hybrid rendering pipeline \cite{Wyman:2018:IDR} that augments diffuse Lambertian shading \cite{Koppal:LR:2014} with ray-traced shadows. 

\subsection{Shading Model}

For deferred shading, rasterization is first performed to generate a G-Buffer, which is a collection of textures containing the data of the nearest fragment per pixel required for lighting computation. Next, light visibility information is obtained via ray tracing for every light per pixel. The G-Buffer and light visibility information are then used to obtain the final pixel colour $I$ as shown.

\begin{equation}
  I = \frac{I_{d}}{\pi} \sum_{i} k_{i} \cdot \saturate{$N \cdot L_{i}$} I_{i}
\end{equation}

$I_{d} / \pi$ refers to the pixel's diffuse bidirectional reflectance distribution function with $I_{d}$ as its material diffuse colour. $N$ denotes its surface normal, and $k_{i}$, $L_{i}$ and $I_{i}$ refer to the relative visibility, direction and intensity of light $i$ respectively in relation to the pixel. For every pixel, we trace a shadow ray from its world position to every light in the scene. If the ray reaches the light without hitting any other object, the light is deemed visible and $k_{i} = 1$, else $k_{i} = 0$. 

The pixel world positions can be obtained from rasterization or ray casting. For rasterization, the server can wait for the client to compute the world positions and send them over or perform rasterization itself to obtain the world positions earlier. To minimize data transfer, both the client and server perform rasterization in our current implementation as it is still relatively fast on both ends. However, while rasterization might have better performance for simpler scenes, modern ray tracing-accelerated hardware can query primary visibility with ray casting even faster than through rasterization. This also streamlines our workflow and avoids repeated computation by omitting rasterization on the server altogether.

In the original non-distributed ray-traced shadow algorithm, the colour contributions of lights are computed and accumulated while looping through every light and tracing rays to it. DHR separates this colour computation and the ray tracing loop so they can be performed on different hardware for better parallelism and performance. As such, the ray tracing process only stores the visibility boolean $k_{i}$ of every light in a visibility bitmap buffer such that every pixel is assigned a compact bitmask representing all the lights in the scene. For instance, if light 2 is visible from a particular pixel, querying the bitmap at its coordinates will produce a bitmask with bit 1 at position 2. The number of bitmaps and size of per-pixel bitmasks can be tuned based on the number of lights in the scene. 

\begin{figure*}
  \includegraphics[width=\linewidth]{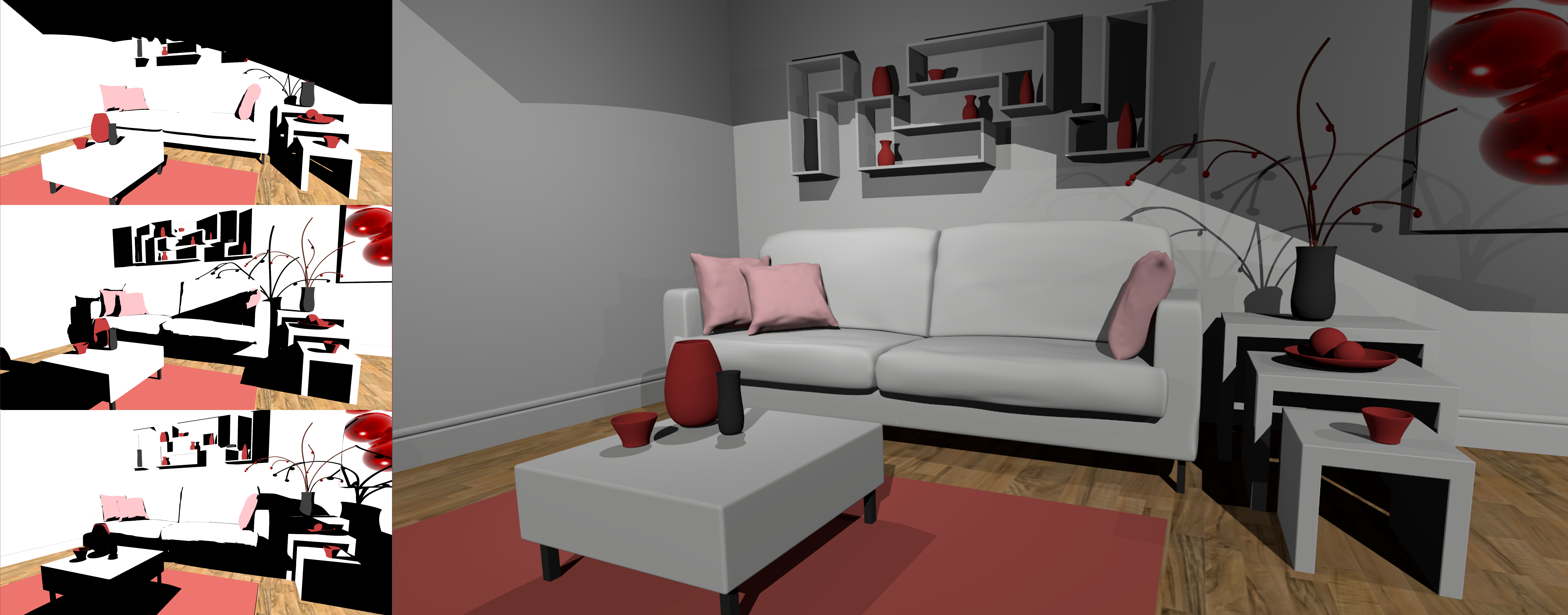}
  \caption{Shading output.}
  \Description{3 images on the left representing each of the 3 lights display visibility information while 1 image on the right shows the final output.}
  \label{fig:visibilitybuffer}
\end{figure*}

We show the hybrid rendering output of the shading model in Figure~\ref{fig:visibilitybuffer} on the scene \textsc{The Modern Living Room} (\href{https://creativecommons.org/licenses/by/3.0/}{CC BY}) by Wig42 \cite{Wig42:2014:MLR} (commonly referred to as \textsc{Pink Room}). The images on the left represent the visualization of the ray-traced visibility buffer encapsulating information for 3 scene lights where the material colour of each pixel is shown if the respective light is not obstructed from its world space position. The right image is the final result of combining the ray-traced shadow information in the visibility buffer with diffuse Lambertian shading at the client.

\subsection{Performance-Accuracy Tradeoff}
\label{sec:tradeoff}

\begin{figure*}
  \centering
  \includegraphics[width=\linewidth]{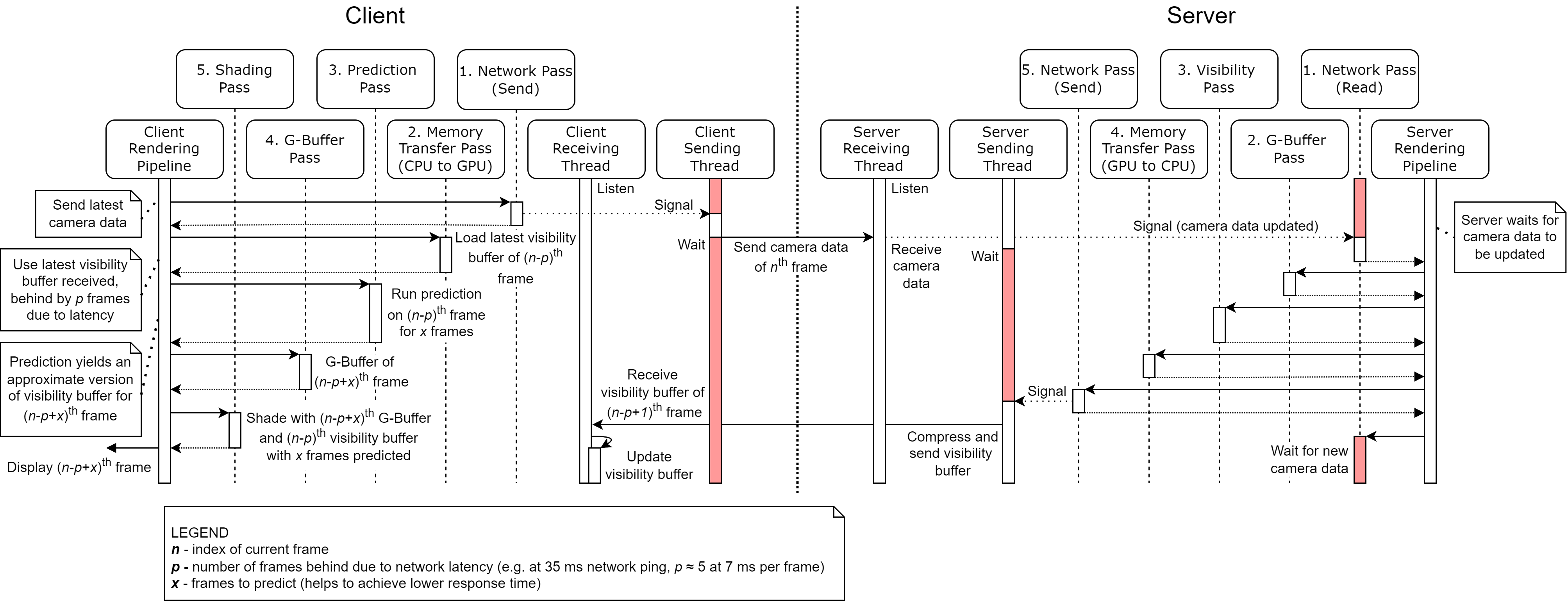}
  \caption{Control flow diagram.}
  \Description{The DHR pipeline includes the G-Buffer, prediction, visibility, shading, network and memory transfer passes.}
  \label{fig:controlflow}
\end{figure*}

Our DHR approach in Figure~\ref{fig:controlflow} adopts the User Datagram Protocol (UDP) for fast data transmission. Although UDP does not retransmit dropped packets, retransmission delays tend to be too costly for interactive applications anyway. Before any rendering is done at the server, the user's scene inputs need to be sent from the client to the server for synchronization. For now, we only handle user-controlled camera movement so the scene inputs consist of camera data (i.e. position, target and up vector). After computing the visibility buffer corresponding to this camera data, the server then compresses it with the LZ4 algorithm which is lossless and efficient on bitmaps. It then sends the compressed buffer and its corresponding frame number to the client for shading computation.

Taking into account network latency, while the client updates the server with the camera data for every frame, the frame it displays to the user is dependent on the latest visibility buffer received from the server. We define response time as the elapsed time from the client sending the camera data to receiving its corresponding visibility buffer. As such, although the client can compute the most updated G-Buffer locally, its latest received visibility buffer will be around $p$ frames behind, where $p$ is the response time divided by the total time taken to render and display one frame. For example, at 90 FPS and a response time of 55 ms (50 ms of network ping and 5 ms of rendering on the server), the client will have access to a visibility buffer of approximately $55 / 11.1 {\approx} 5$ frames behind its G-Buffer. The wait for the correct updated visibility buffer might take too long if the network is slow or even be indefinite if the data is lost as there will be no retransmission from the server.

To strike a balance between performance and accuracy, we provide some allowance for the visibility buffer used such that it can be behind the G-Buffer by a user-defined maximum number of frames $x_{max}$. The value of $x_{max}$ is an artistic choice set by the game developer based on the speed of objects and camera movement in the scene. This is a reasonable approximation as even for dynamic scenes, frames temporally near to each other should be spatially similar, in this case in terms of the occlusion of geometry with respect to scene lights. Nonetheless, applying an outdated visibility buffer as-is with a more recent G-Buffer would lead to shadows not perfectly matching the scene geometry in the final image.

\subsection{Prediction}

Hence, to improve the accuracy of the visibility buffer used, we perform a prediction of the correct visibility buffer based on the scene information with respect to the G-Buffer and offset the visibility buffer to match this scene. In doing so, we minimize the error resulting from scene misalignment between the two buffers.

In Figure~\ref{fig:controlflow}, the client is on its $n^{\text{th}}$ frame with a latest visibility buffer of the $(n - p)^{\text{th}}$ frame, so it can predict $x = min(x_{max}, p)$ frames to obtain an approximation of the $(n - p + x)^{\text{th}}$ visibility buffer and render the more recent $(n - p + x)^{\text{th}}$ frame with the $(n - p + x)^{\text{th}}$ G-Buffer. This allows the client to experience a faster "response time" while keeping the amount of approximation to a pre-defined minimum, maintaining a consistent level of visual quality for the user. For instance, setting $x = 2$ in the example from Subsection~\ref{sec:tradeoff} gives a small lag of around 33 ms or 5 - 2 = 3 frames.

\begin{figure}
  \centering
  \includegraphics[width=\linewidth]{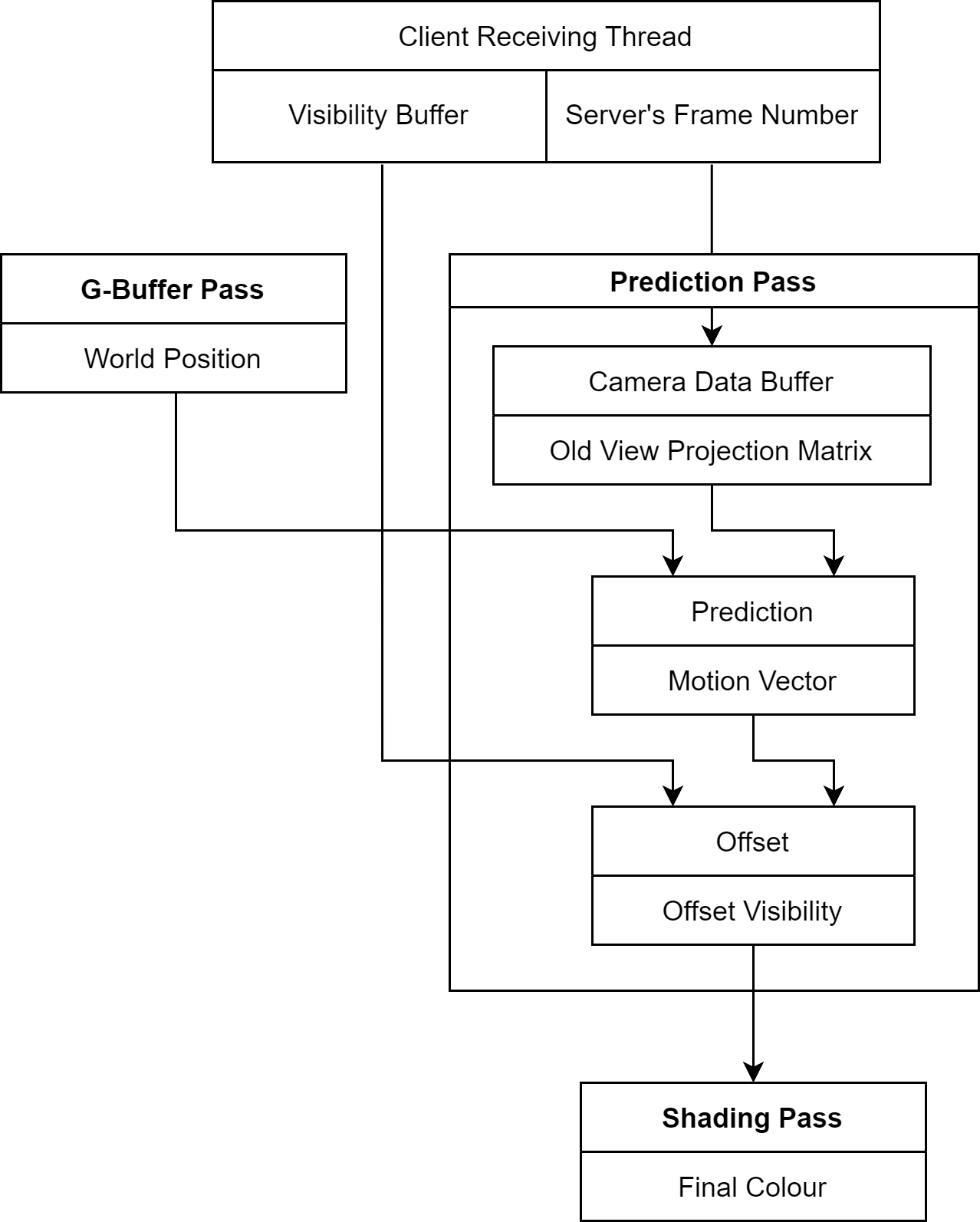}
  \caption{Visibility buffer prediction pipeline.}
  \Description{If the prediction is executed, world space data is used to generate screen-space motion vectors. The motion vectors are then used to offset the received visibility buffer.}
  \label{fig:predictpass}
\end{figure}

For the prediction, the client writes the camera's view projection matrix for every frame in a circular buffer which can store up to the buffer size amount of the latest consecutive frames. As shown in Figure~\ref{fig:predictpass}, it takes the latest visibility buffer along with its corresponding frame number to perform frame prediction. The prediction pass uses the received frame number to retrieve the older camera view projection matrix corresponding to the received visibility buffer. Together with the newer camera view projection matrix of the G-Buffer, the prediction pass calculates a two-dimensional motion vector for every pixel. These motion vectors store the difference between the screen space position of the same world space points as seen by the old camera and the new camera. Finally, each pixel's visibility information in the received visibility buffer will be offset by the individual motion vectors, forming a new visibility buffer that will be used for the shading of the scene.

We do not know the visibility information of scene points revealed by camera movement, so their corresponding pixels are taken to be fully illuminated in relation to all scene lights. Doing so helps to prevent adding false shadows to the scene if there is camera movement. However, there are prominent borders of illuminated pixels along the edges of the output frame in the direction of camera movement as the offset positions given by the motion vectors fall outside the boundary of the original visibility buffer. We minimize this error at a small performance cost by working with slightly larger textures so some visibility information is available past the borders of the frame. For example, taking a display resolution of 1920 $\times$ 1080, we work with a larger 1984 $\times$ 1116 visibility buffer texture with an additional 64 $\times$ 36 pixels. This visibility buffer size is another variable that the game developer can set depending on the interactive performance requirement of the application and the speed of camera movement. The focal length of the camera is adjusted to maintain the same displayed region of the scene for different numbers of additional pixels to the display resolution.

%% file: evaluation.tex
\section{Evaluation}

The results of our DHR implementation are measured on a client as well as a server with the ray tracing-accelerated NVIDIA GeForce RTX 2080 GPU. Both the client and server have an Intel Core i7-7700K central processing unit (CPU) and 16 GB random-access memory (RAM). Our qualitative measurements are taken over a 14 second and 50 second camera animation sequence for the scenes \textsc{Pink Room} (PR) and \textsc{UE4 Sun Temple} (\href{https://creativecommons.org/licenses/by-nc-sa/4.0/}{CC BY-NC-SA}) by \citet{EpicGames:2017:UES} (ST) respectively at a display resolution of 1920 $\times$ 1080.

\subsection{Graphics Quality}

\begin{figure*}
    \centering
    \subcaptionbox{No prediction}{
        \includegraphics[width=0.238\linewidth]{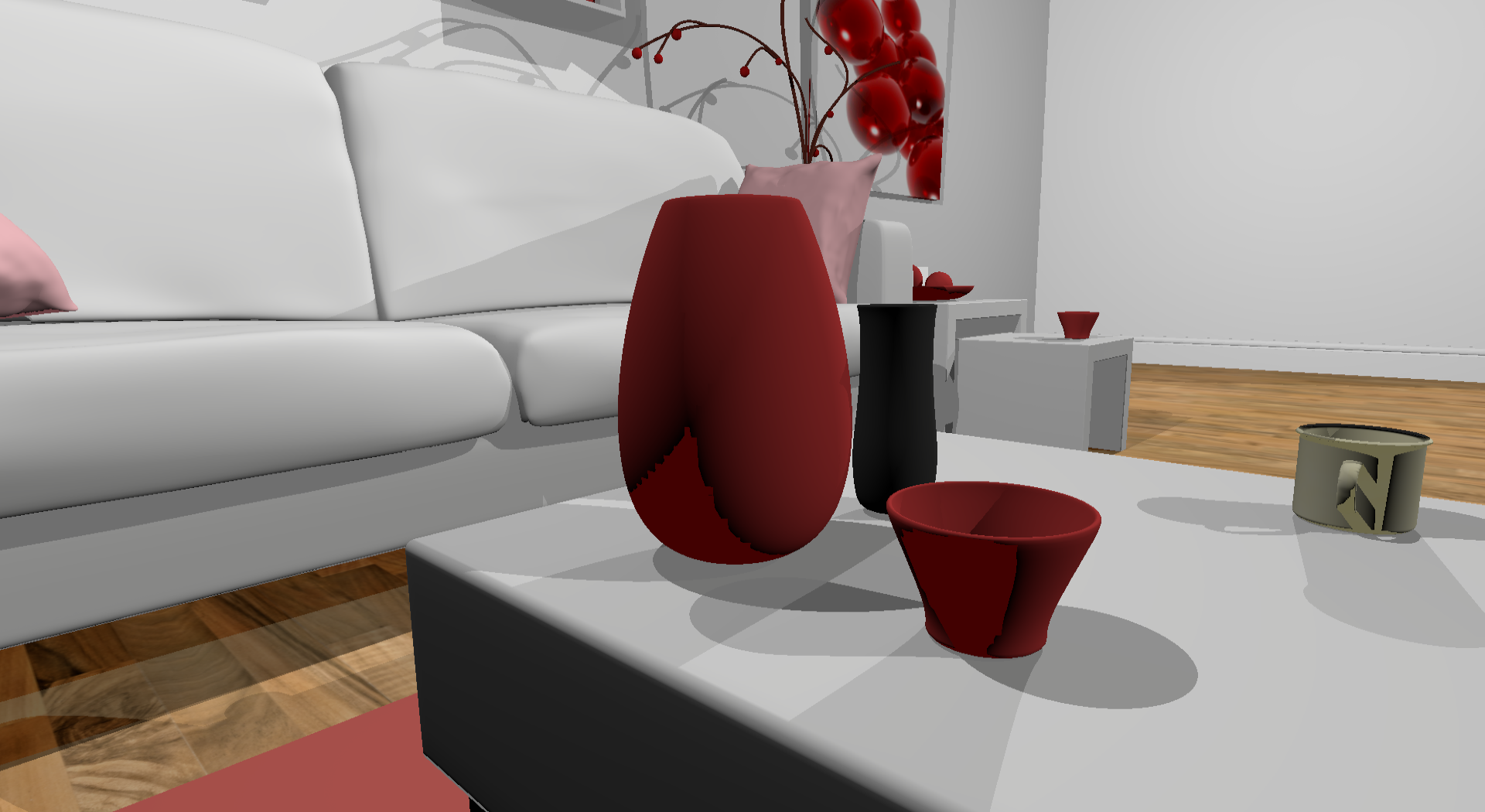}
    }
    \subcaptionbox{$x = 1$}{
        \includegraphics[width=0.238\linewidth]{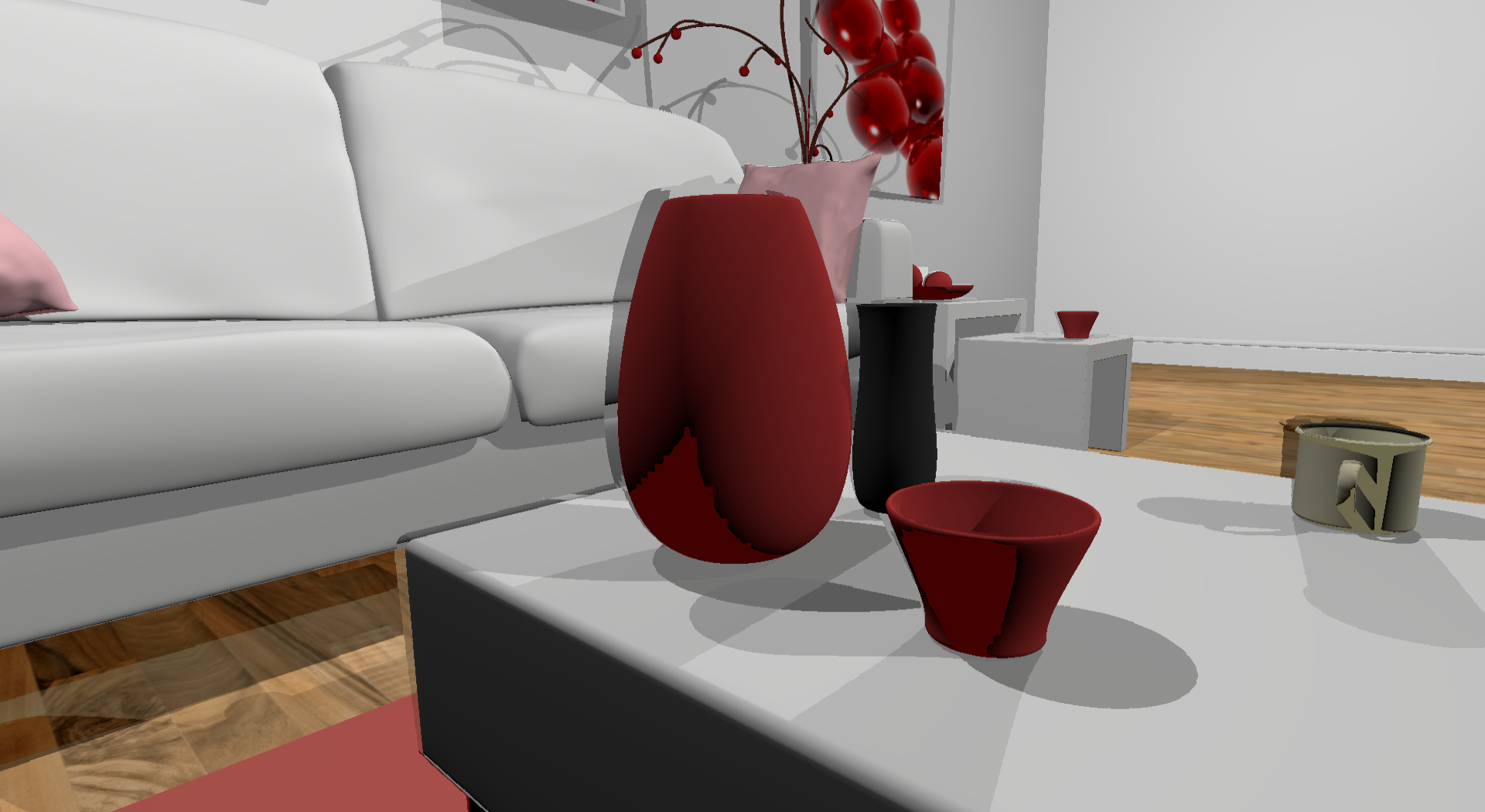}
    }
    \subcaptionbox{$x = 5$}{
        \includegraphics[width=0.238\linewidth]{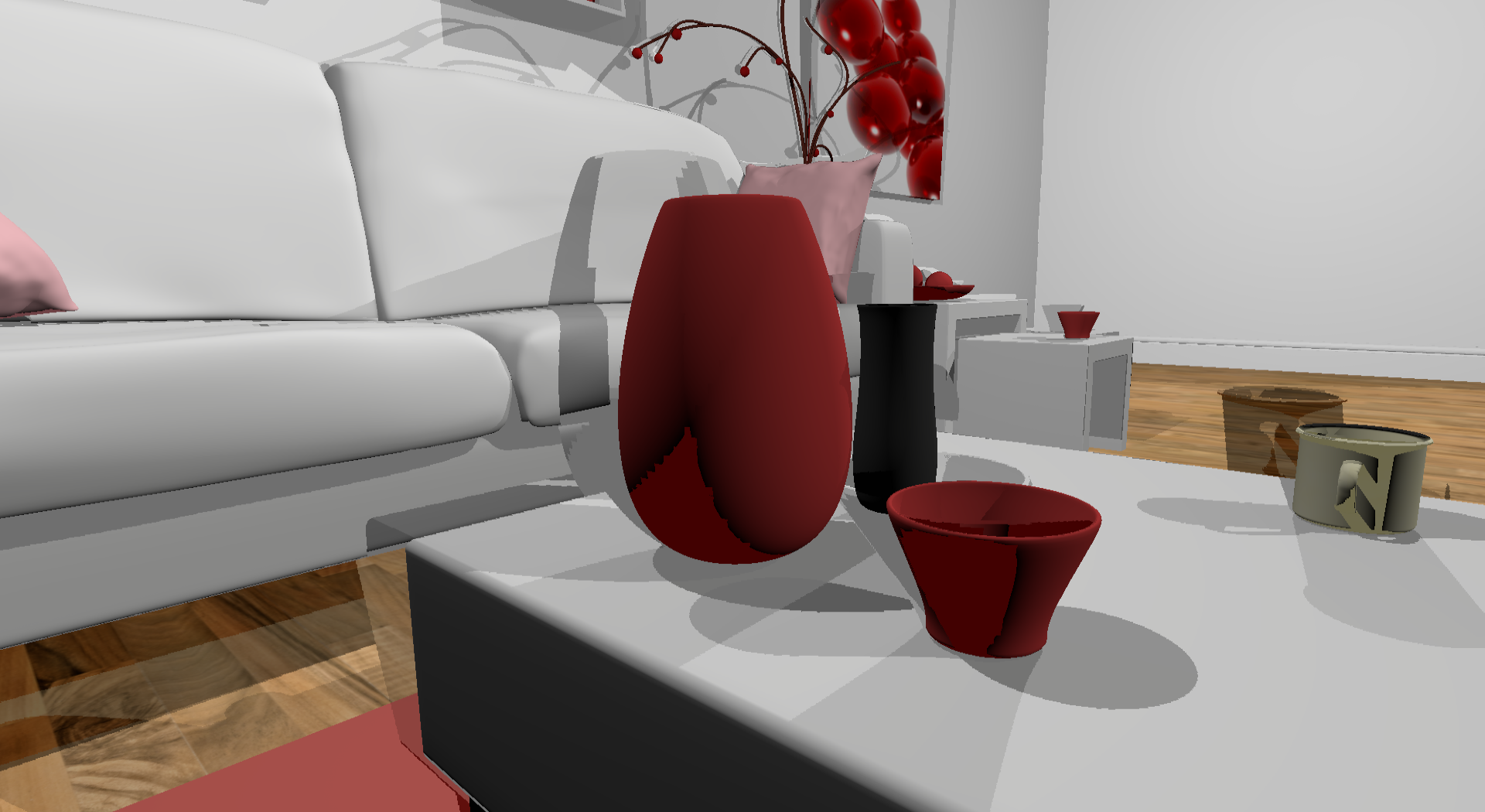}
    }
    \subcaptionbox{$x = 10$}{
        \includegraphics[width=0.238\linewidth]{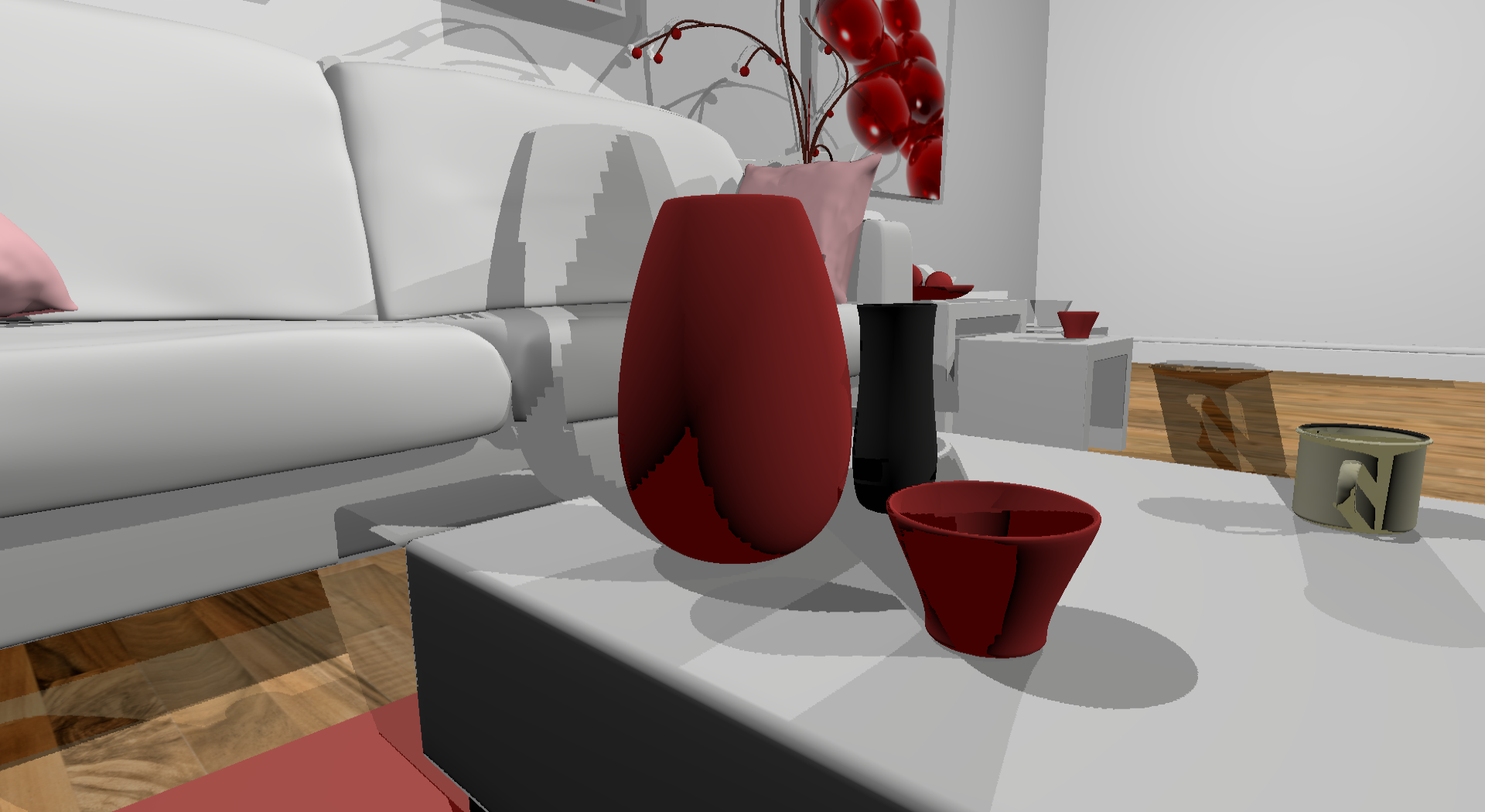}
    }
    \Description{Pink Room, medium shot of coffee table with sofa in background}
    \caption{PR (454,505 vertices).}
    \label{fig:pinkroom}
\end{figure*}

\begin{figure*}
    \centering
    \subcaptionbox{No prediction}{
        \includegraphics[width=0.238\linewidth]{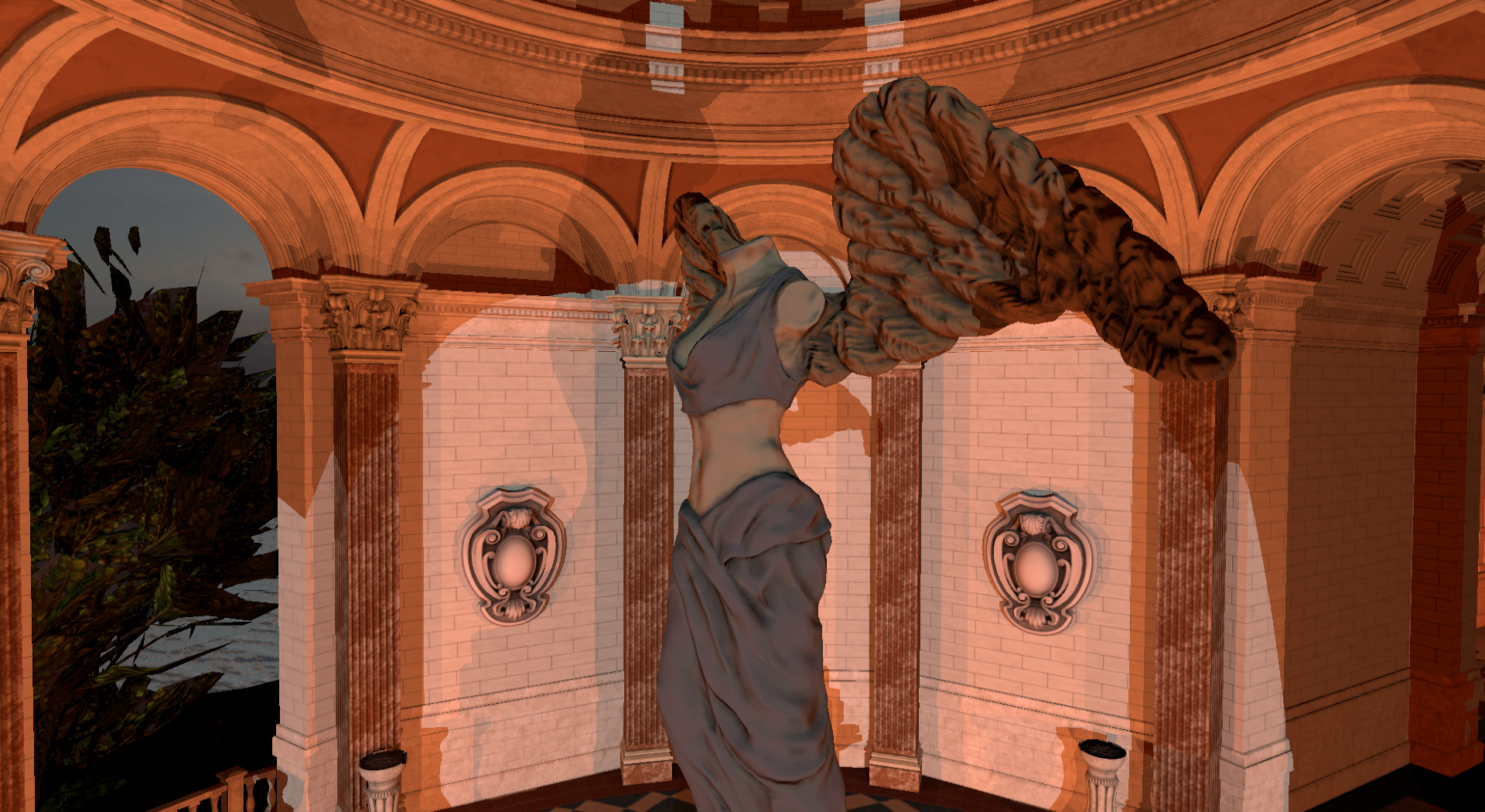}
    }
    \subcaptionbox{$x = 1$}{
        \includegraphics[width=0.238\linewidth]{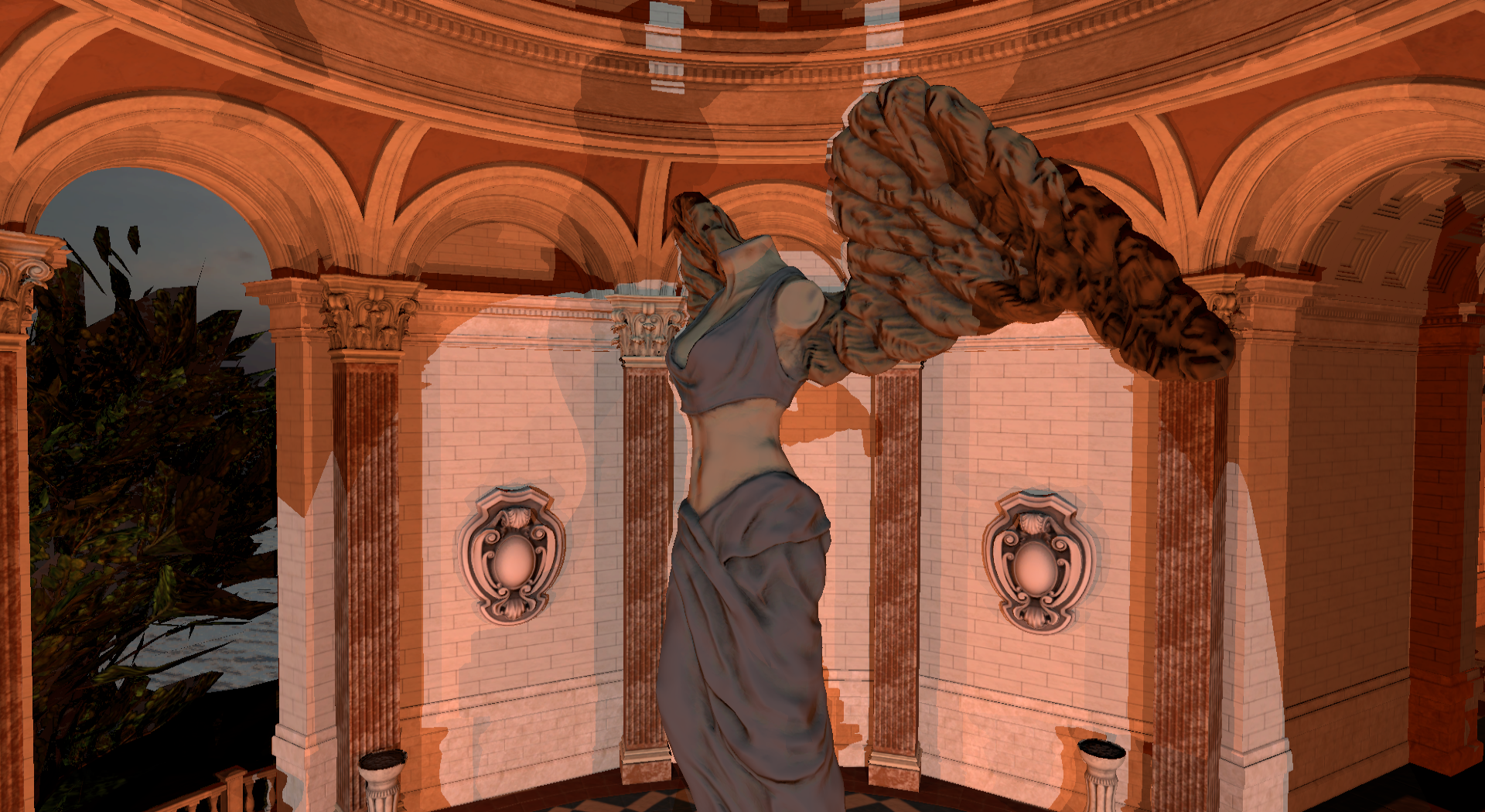}
    }
    \subcaptionbox{$x = 5$}{
        \includegraphics[width=0.238\linewidth]{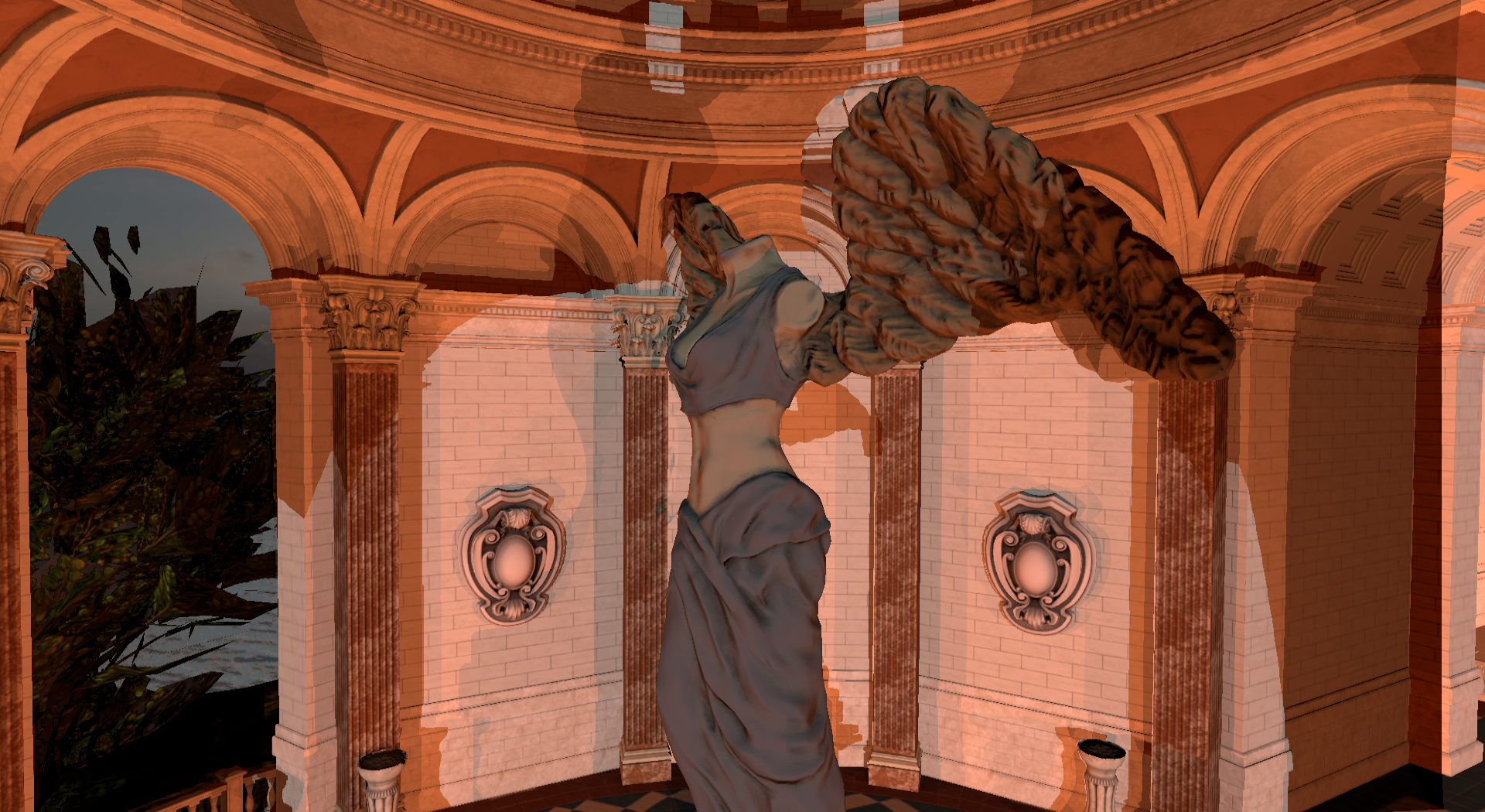}
    }
    \subcaptionbox{$x = 10$}{
        \includegraphics[width=0.238\linewidth]{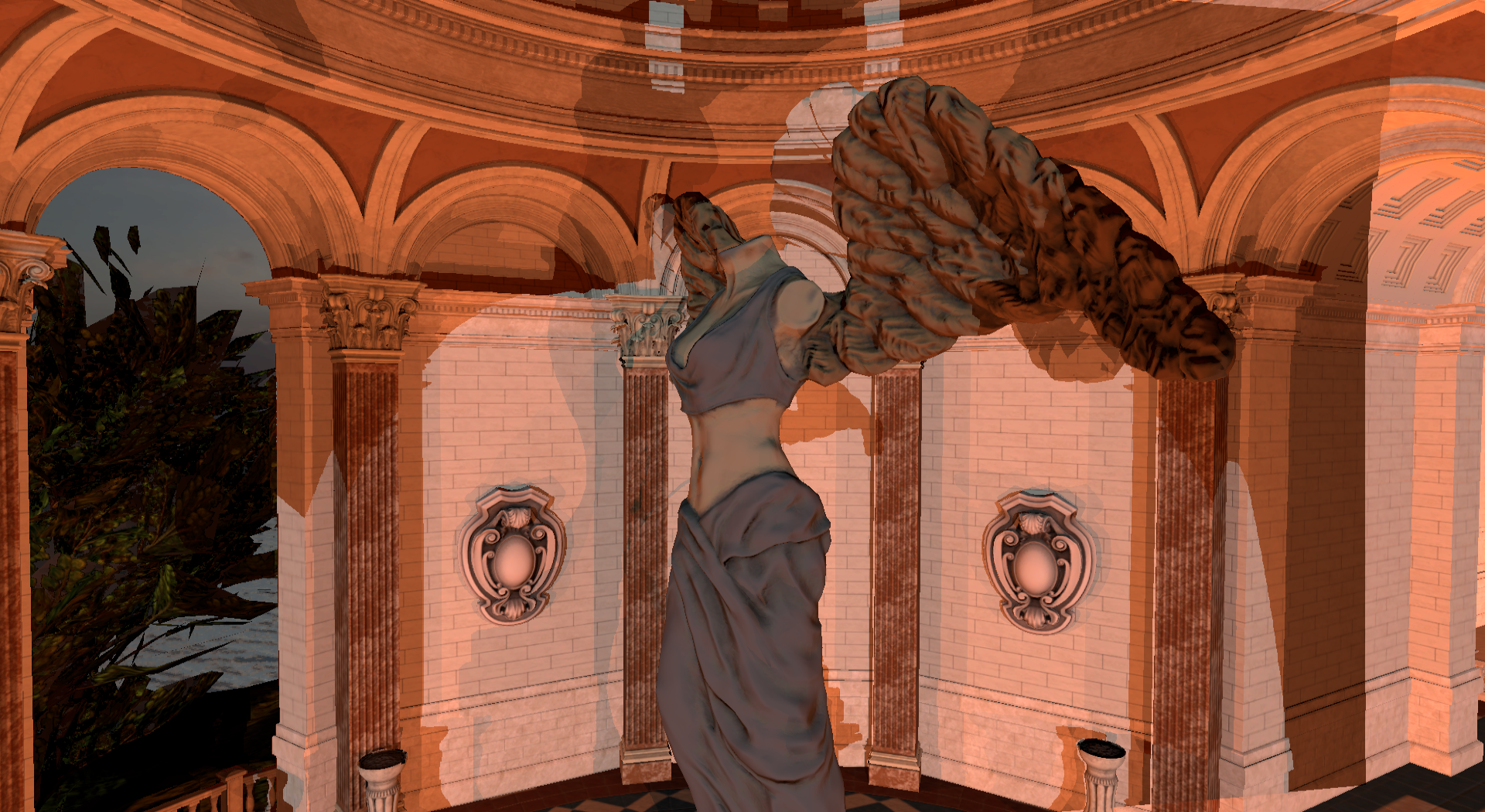}
    }
    \Description{Sun Temple, medium shot of angel casting a shadow on the background wall}
    \caption{ST (1,641,711 vertices).}
    \label{fig:suntemple}
\end{figure*}

\begin{figure*}
    \centering
    \includegraphics[width=0.242\linewidth]{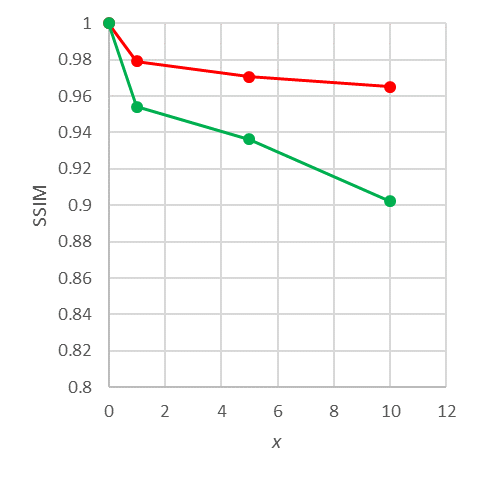}
    \includegraphics[width=0.242\linewidth]{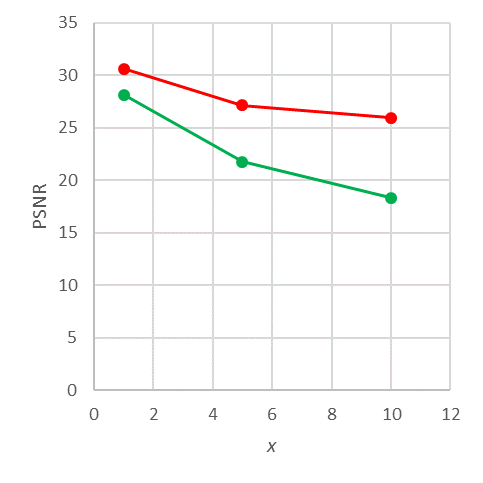}
    \includegraphics[height=0.242\linewidth]{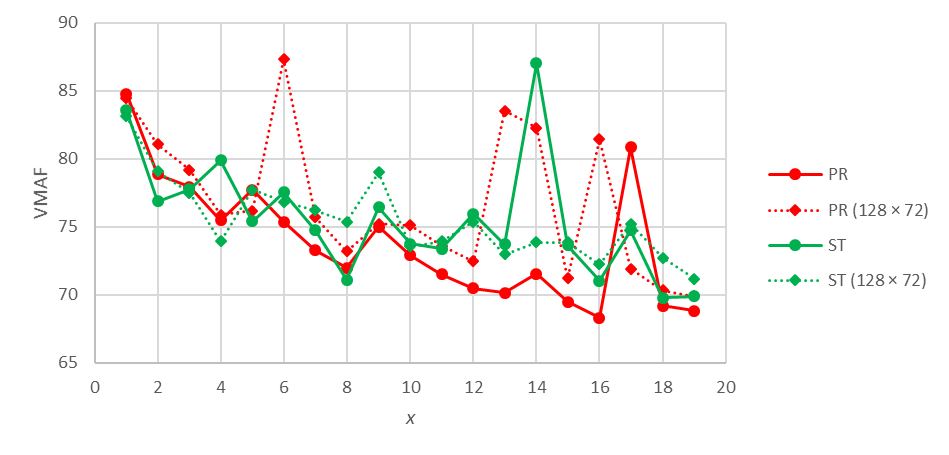}
    \Description{SSIM, PSNR and VMAF values decrease as the number of frames predicted increases.}
    \caption{Objective visual metrics.}
    \label{fig:metrics}
\end{figure*}

\begin{figure*}
  \centering
  \includegraphics[height=0.373\linewidth]{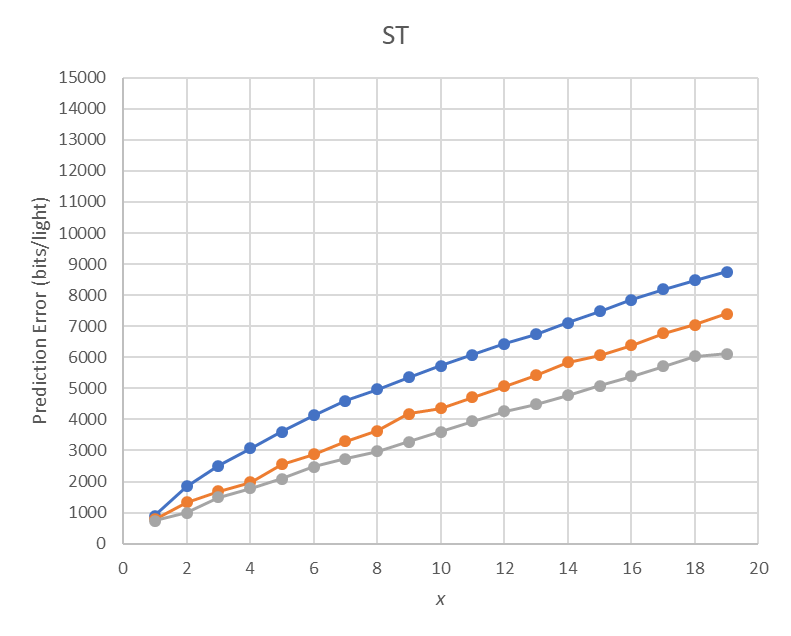}
  \includegraphics[height=0.373\linewidth]{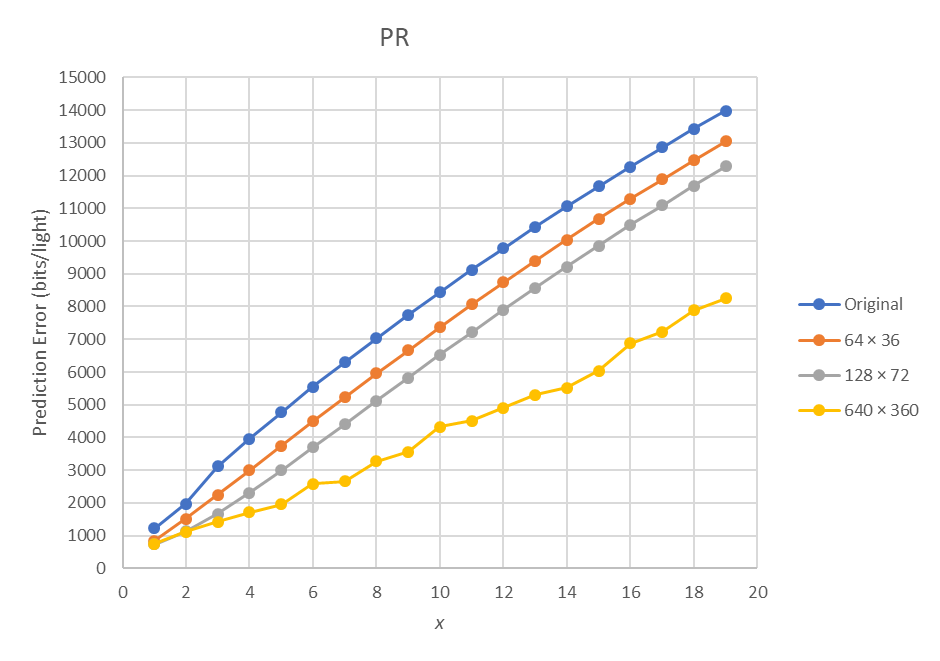}
  \caption{Prediction error.}
  \Description{Error is large when the visibility buffer is small and the number of frames predicted is large.}
  \label{fig:bitwiseerror}
\end{figure*}

We show the result of our visibility buffer prediction on the final output with camera movement. As seen in Table~\ref{tab:predictdemo}, our prediction pass improves the accuracy of the shadows. With an additional 128 $\times$ 72 pixels on the visibility buffer, we obtain Figures~\ref{fig:pinkroom} and ~\ref{fig:suntemple} and their corresponding objective visual metrics measurements as shown in Figure~\ref{fig:metrics}. The visual quality of the output decreases as more frames $x$ are predicted for the same frame $n$. This is because as $x$ increases while $n$ remains constant, $n - p$ decreases which means that an older visibility buffer is used for the prediction, resulting in a higher bitwise prediction error as shown in Figure~\ref{fig:bitwiseerror}. From our preliminary testing with some users, $x$ is best capped at 4 or 5.

\begin{table*}
\begin{center}
    \caption{Prediction output.}
    \label{tab:predictdemo}
\begin{tabular}{|c|c|c|c|}
    \hline
    \textbf{Delay (ms)} & 
    \textbf{0} & 
    \textbf{100} & 
    \textbf{200} \\ \hline
    Before Prediction & 
    \multirow{-8.2}{*}{\includegraphics[width=0.398\linewidth]{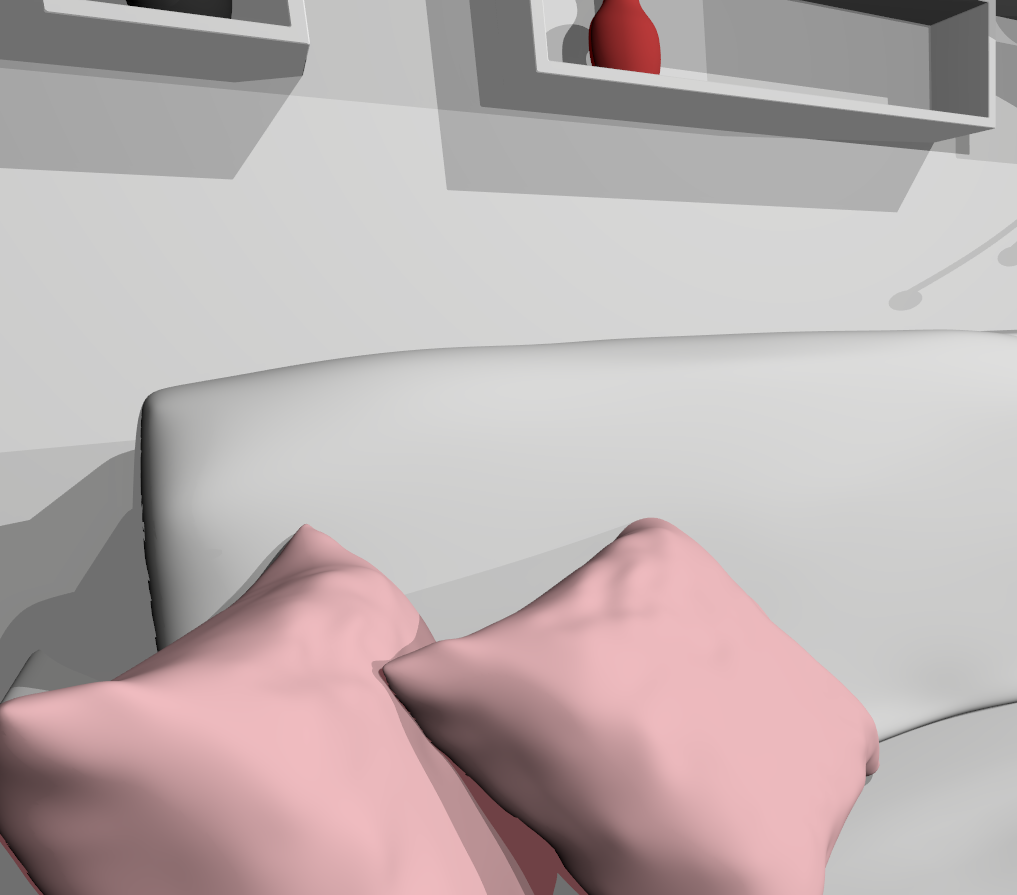}} &
    \includegraphics[width=0.195\linewidth]{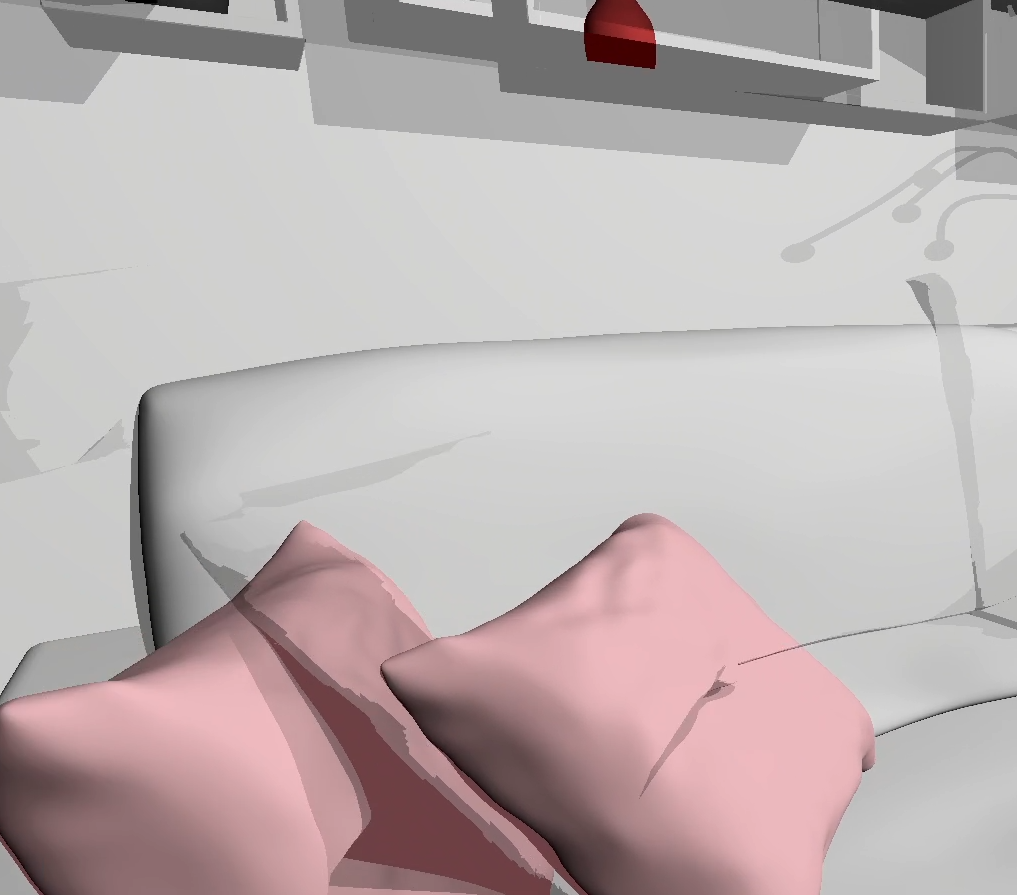} &
    \includegraphics[width=0.195\linewidth]{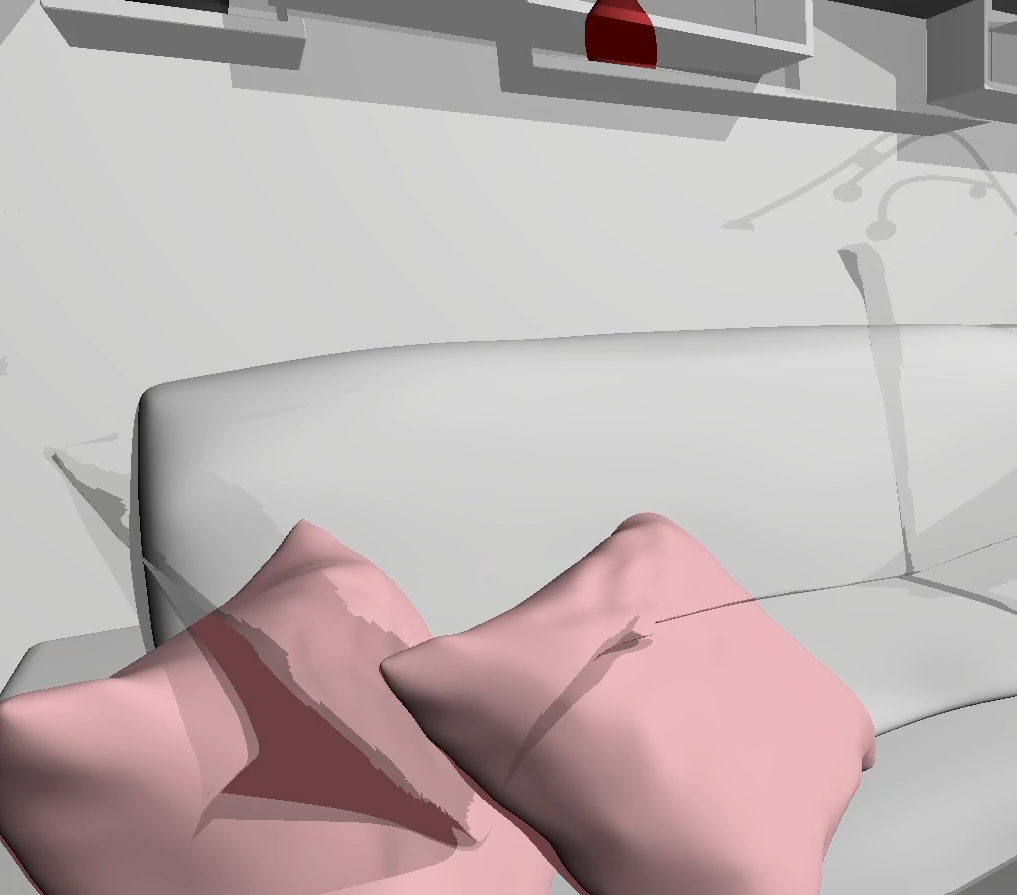} \\ \cline{1-1} \cline{3-4}
    After Prediction & & 
    \includegraphics[width=0.195\linewidth]{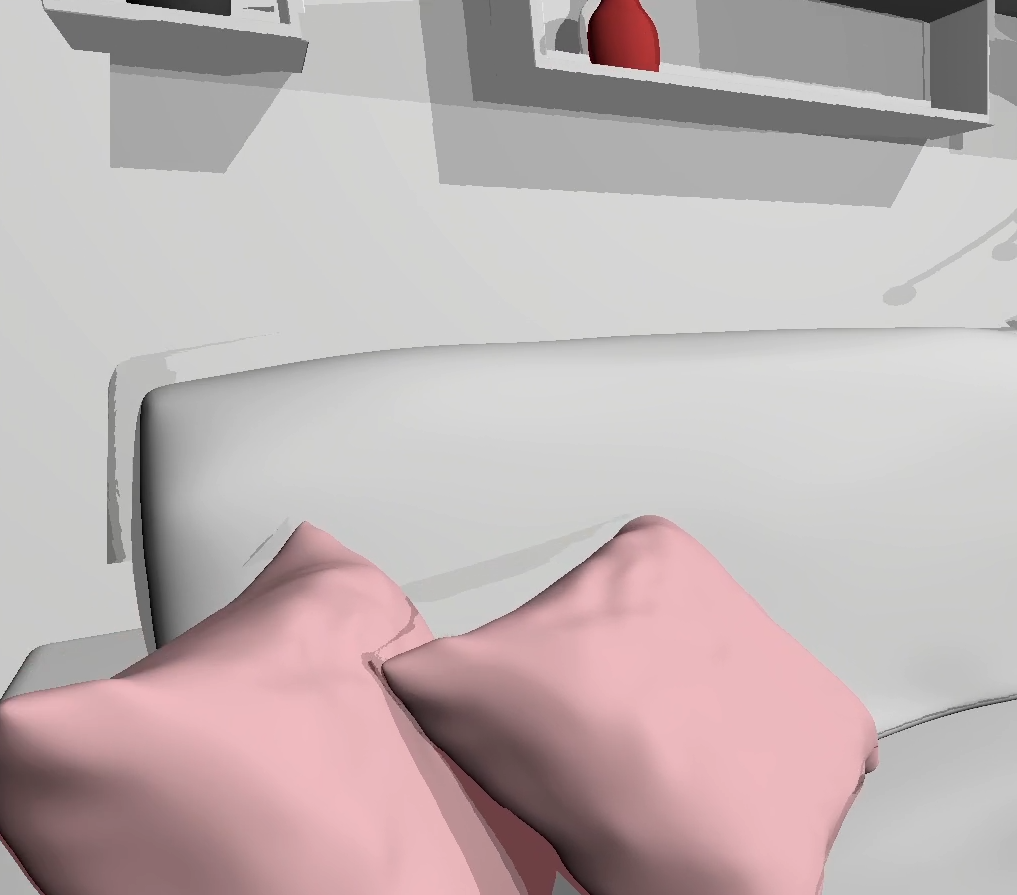} & 
    \includegraphics[width=0.195\linewidth]{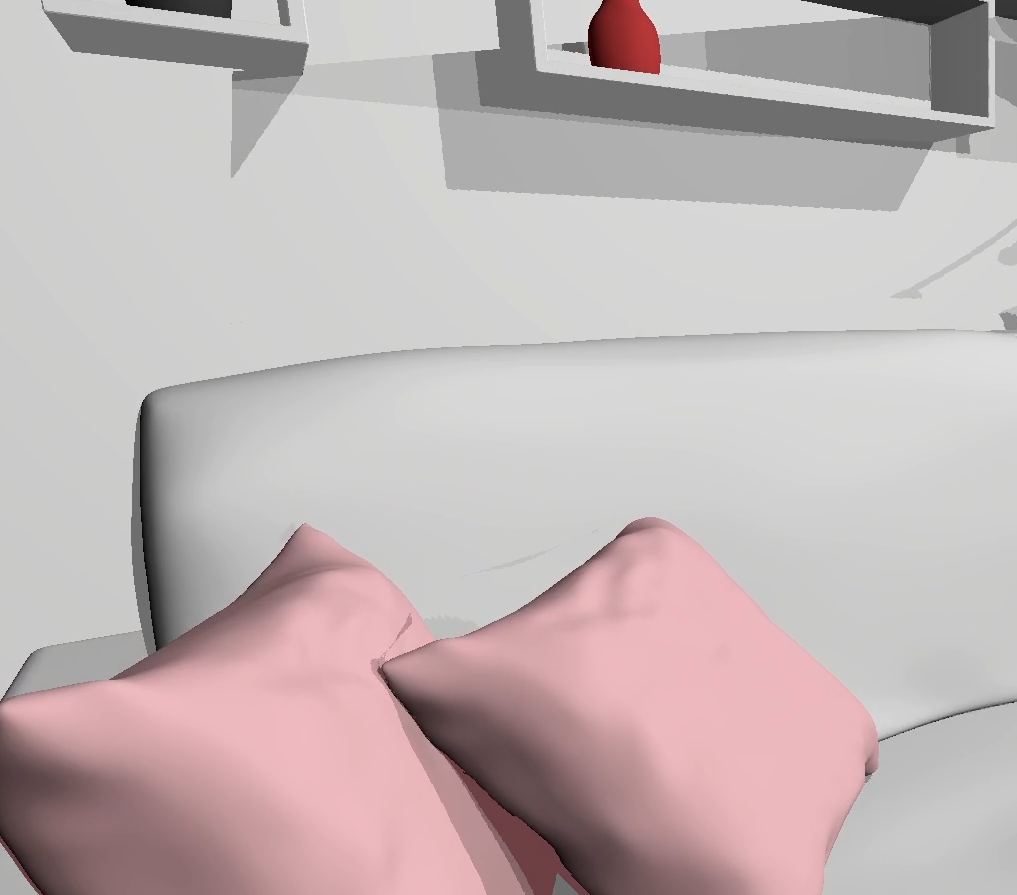} \\ \hline
\end{tabular}
\end{center}
\end{table*}

Under the fixed laboratory testing environment used for our experiments, the Structural Similarity Index Measure (SSIM), Peak Signal-to-Noise Ratio (PSNR) and Video Multimethod Assessment Fusion (VMAF) scores obtained do not vary across repeated experiments. The VMAF data also does not highlight the impact of the visibility buffer size on the accuracy of the prediction. As such, there is room for improvement in finding a better metric to measure the human perception of error arising from the misalignment of the visibility buffer. Nonetheless, there is a general downward trend at a low number of frames predicted, showing that human perception of the prediction errors can be felt. However, there are anomalies when more frames are predicted. This could be because VMAF is mostly trained to identify and evaluate the effects of artifacts and blurriness arising from lossy encoding. While the artifacts caused by the prediction error may resemble such artifacts when small, VMAF may not be able to accurately evaluate the video when the prediction error and hence image distortion becomes significant. Regardless, even amongst the anomalies, we can observe a downward trend.

Figure~\ref{fig:bitwiseerror} shows the average prediction error per light measured using the bitwise comparison of the actual visibility buffer with the predicted version. We see that increasing the size of the visibility buffer does lower the prediction error by roughly a constant amount, which can only arise from minimizing errors at the edges of the frame since increasing buffer size does not affect anything else. This has a large impact and is visually very noticeable at lower values of frames predicted, where the error can be almost halved with the use of 128 $\times$ 72 additional pixels. We included a result of using an unreasonably large buffer of 2560 $\times$ 1440 pixels (i.e. 640 $\times$ 360 additional pixels) on PR to show that a large portion of the error arises from the edges of the frame. The same result was not measured on the larger ST due to the difficulty of maintaining the frame rate while performing the error calculation. 

The prediction error calculations used samples recorded at a fixed frame rate of 60 FPS as an industry standard frame rate for real-time interactive applications like games. VMAF samples were recorded at a lower 30 FPS to reduce deviation from VMAF’s 24 FPS frame rate recommendation. A discrepancy from the recommended value would not affect the reading significantly, but it will be interpreted as a 24 FPS video so the motion feature of the metric will generate lower scores and cause the overall VMAF score to be slightly lower.

\begin{figure}
    \centering
    \subcaptionbox{$x = 1$}{
        \includegraphics[width=0.47\linewidth]{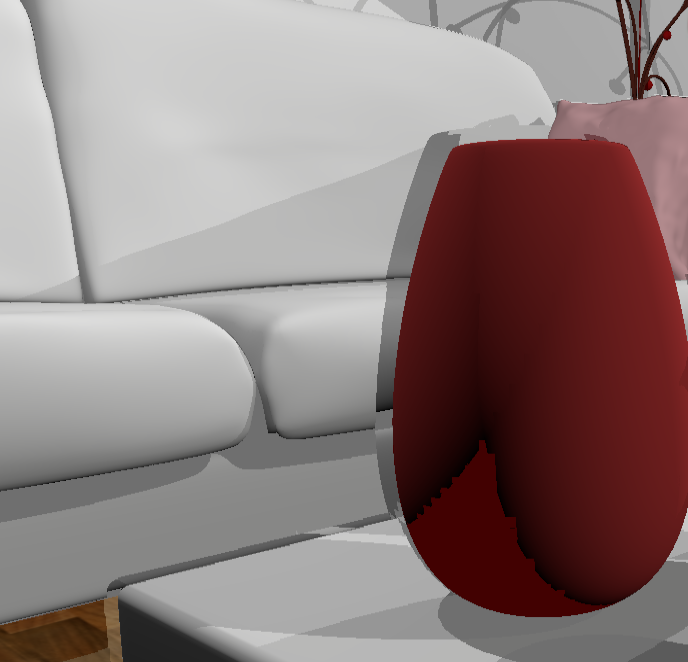}
    }
    \subcaptionbox{$x = 10$}{
        \includegraphics[width=0.47\linewidth]{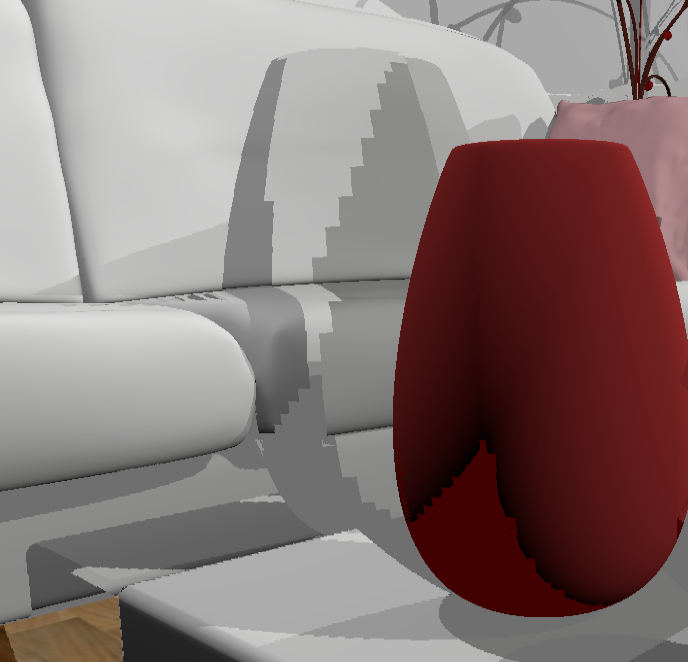}
    }
    \Description{Pink Room, close-up shot of red vase on coffee table}
    \caption{Figure~\ref{fig:pinkroom} (PR) zoomed in.}
    \label{fig:pinkroom-vase}
\end{figure}

\begin{figure}
    \centering
    \subcaptionbox{$x = 1$}{
        \includegraphics[width=0.47\linewidth]{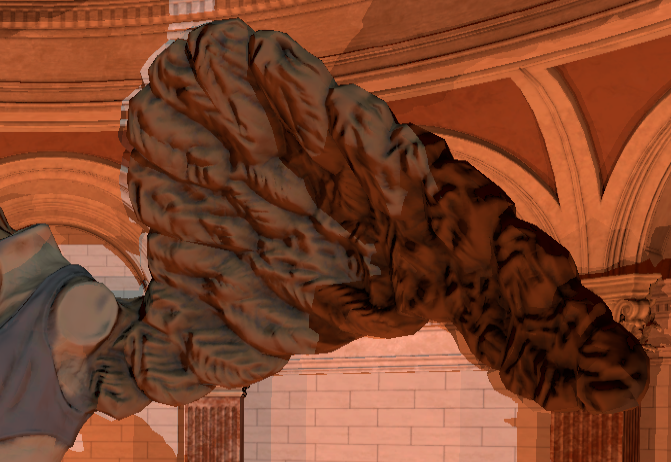}
    }
    \subcaptionbox{$x = 10$}{
        \includegraphics[width=0.47\linewidth]{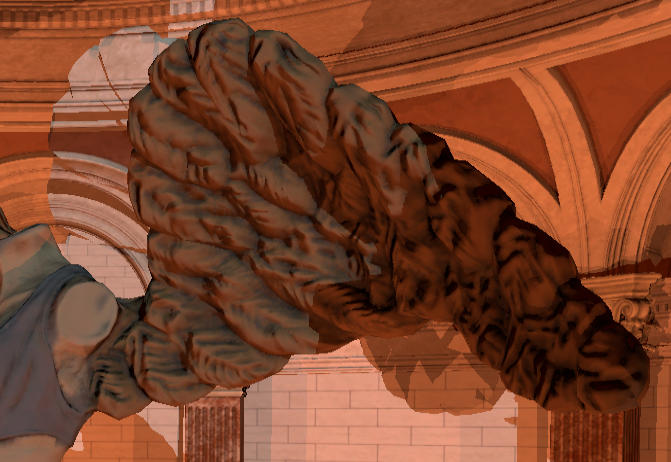}
    }
    \Description{Sun Temple, close-up shot of right angel wing}
    \caption{Figure~\ref{fig:suntemple} (ST) zoomed in.}
    \label{fig:suntemple-wing}
\end{figure}

Qualitatively, the shadows for both scene setups generally remain consistent except for those cast by foreground geometry onto background objects. For example, the moving shadows of the red vase in Figure~\ref{fig:pinkroom-vase} and the angel statue in Figure~\ref{fig:suntemple-wing} with different numbers of frames predicted show how the visibility information of fragments in the interior of the frame revealed by camera movement cannot be reliably predicted by offsetting the visibility buffer of previous frames which do not contain this information in the first place. As for pixels at the edges of the frame, Figure~\ref{fig:pinkroom} does not have the issue of illuminated pixel borders like Figure~\ref{fig:suntemple}, implying that the ST scene setup here needs to be rendered with a larger visibility buffer texture. Additionally, in Figure~\ref{fig:pinkroom-vase}, we see jaggies along the shadow edge as it is enlarged from the original visibility buffer to the offset one due to camera movement, so it now occupies more pixels. This results in multiple nearby pixels querying the original visibility buffer at the same coordinates which can be avoided with linear sampling instead of point sampling. We can also see that in Figure~\ref{fig:pinkroom}, the shadow on the floor of the beige-coloured mug which is moving in PR is not predicted accurately as we currently do not handle dynamic scene geometry in our visibility buffer prediction.

\subsection{Performance}

In our setup, the client and server are connected via Ethernet over a local network and are physically located very close to each other (i.e. in the same lab). As such, the latency between them is less than 1 ms. Hence, we add artificial network ping via network condition simulator \href{http://jagt.github.io/clumsy/}{\emph{clumsy}} and measure the frame rate and response time of our approach on different network latency conditions. The results in Table~\ref{tab:fps} are as expected as they show that frame rate decreases with increased geometric complexity of the scene and additional pixels of the visibility buffer. In general, we achieve interactive frame rates for our scene setups even when rendering large visibility buffers.

\begin{table}
\begin{center}
  \caption{Frame rate.}
  \label{tab:fps}
\begin{tabular}{|c|c|c|}
 \hline
 \textbf{Additional Pixels} & \textbf{PR} & \textbf{ST} \\ \hline
 0 & 154 & 133 \\ \hline
 64 $\times$ 36 & 148 & 127  \\ \hline
 128 $\times$ 72 & 142 & 125 \\ \hline
\end{tabular}
\end{center}
\end{table}

\begin{figure}[!ht]
  \centering
  \includegraphics[width=\linewidth]{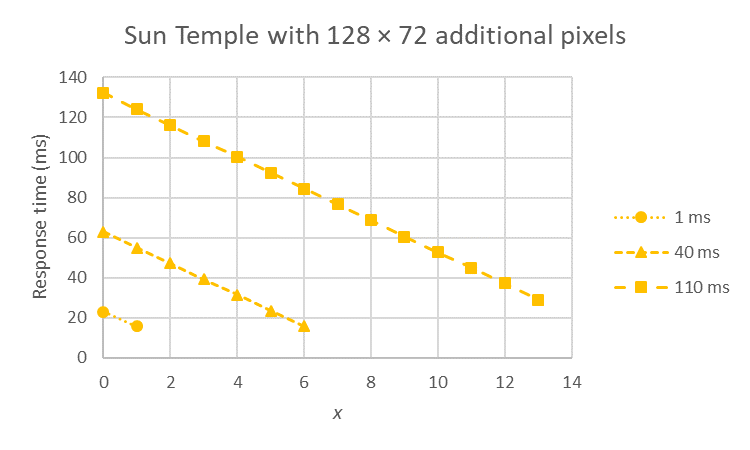}
  \includegraphics[width=\linewidth]{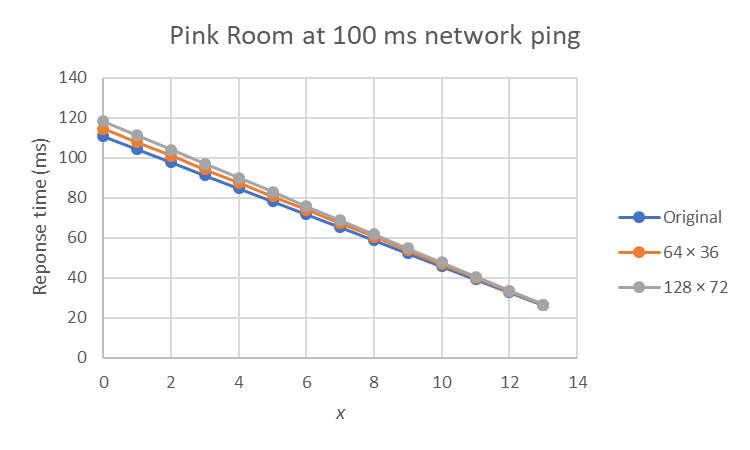}
  \caption{Response time.}
  \Description{Response time decreases as the number of frames predicted increases}
  \label{fig:latency}
\end{figure}

As shown in Figure~\ref{fig:latency}, the response time is fast for low network ping values. Initially at no frames predicted, the response time is higher with a larger buffer as it takes slightly longer for the server to transfer between device and host memory, render, compress and send the texture. Response time then decreases as expected with more frames predicted. We also note that the response time experienced when using different visibility buffer sizes converges as seen in the bottom graph. This is because every additional frame predicted reduces the response time more for larger visibility buffers which take more time to generate and send to the client.

%% file: futurework.tex
\section{Future Work}

To improve the accuracy of our output, one possible modification to the pipeline could be to perform server-side prediction \citep{Bhojan:2017:CTL} in addition to the current client-side prediction. The server would render a few different visibility buffers based on cases of potential camera movement. These visibility buffers would be almost identical and contain much duplication of data, so they should be able to be highly compressed before they are sent to the client. The client will then select the visibility buffer that corresponds most closely to the actual camera motion and use that for the client-side prediction.

Currently, if the server cannot be reached at all, the client erroneously performs lighting computation with an empty visibility buffer. However, in extreme cases of poor network conditions, we are intending to fall back on local rendering with just rasterization as a last resort. This enables us to generate interactive output for the user even if the connection is too slow or lost, or if the remote server is down. We would be able to maintain a baseline of temporal coherence for the user in such situations by switching to local rendering dynamically when network latency values exceed a certain threshold, which could be a variable we expose to developers for trading-off between visual quality and performance.

Performing rasterization on the client while ray tracing on the server in parallel can give an overall improvement in performance as compared to pure remote rendering when enabled by fast 5G networks and low latencies to reach 5G edge nodes \citep{Bhojan:2014:GMM}. Hence, we intend to test our implementation in a simulated 5G edge computing environment to leverage the fast network technology.

We are also looking to expand the scope of scene inputs to accommodate dynamic changes in scene geometry in addition to camera movement. Besides affecting the scene synchronization between the client and server, this upgrade will also impact the prediction pass. On top of the camera's view projection matrix, the circular buffer would also need to hold the information of moving objects for every frame, which would have to be taken into account when calculating the screen space motion vector of each pixel.

The compression quality of visibility buffer packets of the same frame can also be varied to save on bandwidth. The user's attention is more likely to focus on certain regions in the frame so \citet{Babaei:2017:SGA} maps out predictions of such regions and allocates more bitrate accordingly, while \citet{Illahi:2020:CGF} employs a gaze tracker device to determine the exact spot the user is looking at.

In the future, we hope to incorporate multiple complex lighting and camera effects into a single hybrid rendering pipeline as part of an adaptive hybrid system that can dynamically adjust parameters for the different effects to achieve the maximum visual quality while maintaining interactive frame rates, being adaptive to changing scene properties and hardware configurations. We also hope for this pipeline to adopt a DHR approach for ray tracing-incorporated graphics in standalone XR devices as part of metaverse experiences.

%% file: conclusion.tex
\section{Conclusion}

We present the Distributed Hybrid Rendering (DHR) approach for standalone extended reality (XR) devices through the design and evaluation of a simple prototype which generates ray-traced shadows without compromising the interactive frame rates required for immersive metaverse experiences. Our technique is elastic to network conditions, dynamically adjusting the amount of visibility buffer approximation to maximize the visual quality of the output while maintaining interactive frame rates. 